\documentclass{amsart}
\usepackage{amsmath}
\usepackage{latexsym}
\usepackage{amsfonts}
\usepackage{amssymb}
\usepackage{graphicx}
\newtheorem{theorem}{Theorem}

\newtheorem{lemma}[theorem]{Lemma}

\begin{document}

\author{Brian C. Hall}
\address{Department of Mathematics\\
University of Notre Dame \\
Notre Dame, IN 46556 U.S.A.}
\email{bhall@nd.edu}
\title{Bounds on the Segal-Bargmann transform of $L^{p}$ functions}
\date{February, 2001}
\begin{abstract}
This paper gives necessary conditions and slightly stronger sufficient
conditions for a holomorphic function to be the Segal-Bargmann transform of a
function in $L^{p}(\mathbb{R}^{d},\rho),$ where $\rho$ is a Gaussian measure.
The proof relies on a family of inversion formulas for the Segal-Bargmann
transform, which can be ``tuned'' to give the best estimates for a given value
of $p.$ I also give a single necessary-and-sufficient condition for a
holomorphic function to be the transform of a function $f$ such that any
derivative of $f$ multiplied by any polynomial is in $L^{p}(\mathbb{R}%
^{d},\rho).$ Finally I give some weaker but dimension-independent conditions.
\end{abstract}

\maketitle

\section{Introduction}

I consider the Segal-Bargmann transform for $\mathbb{R}^{d}$ in the following
form. Let $\rho$ denote the standard Gaussian density on $\mathbb{R}^{d},$
namely,
\[
\rho\left(  x\right)  =\left(  2\pi\right)  ^{-d/2}e^{-x^{2}/2}\,dx,
\]
where here and throughout the paper, $x^{2}=x_{1}^{2}+\cdots+x_{d}^{2}.$ We
will then consider the associated Gaussian measure $\rho\left(  x\right)
\,dx.$ We will let $L^{p}(\mathbb{R}^{d},\rho)$ stand for $L^{p}%
(\mathbb{R}^{d},\rho\left(  x\right)  \,dx).$ Note that $\rho\left(  x\right)
$ has an entire analytic continuation to $\mathbb{C}^{d}$.

Now consider $f\in L^{p}(\mathbb{R}^{d},\rho),$ with $1<p<\infty.$ Define the
\textbf{Segal-Bargmann transform} $Sf$ of $f$ by
\begin{align}
Sf\left(  z\right)   &  =\int_{\mathbb{R}^{d}}\rho\left(  z-x\right)  f\left(
x\right)  \,dx,\nonumber\\
&  =(2\pi)^{-d/2}\int_{\mathbb{R}^{d}}e^{-(z-x)^{2}/2}f(x)\,dx,\quad
z\in\mathbb{C}^{d}.\label{sdef1}%
\end{align}
This may also be written as
\begin{equation}
Sf\left(  z\right)  =e^{-z^{2}/2}\int_{\mathbb{R}^{d}}e^{z\cdot x}f\left(
x\right)  \rho\left(  x\right)  \,dx,\quad z\in\mathbb{C}^{d}.\label{sdef2}%
\end{equation}
Since the function $e^{z\cdot x}$ is in $L^{p^{\prime}}(\mathbb{R}^{d},\rho)$
(with $p^{\prime}$ the conjugate exponent to $p$), the integral is absolutely
convergent for all $f\in L^{p}(\mathbb{R}^{d},\rho).$ Using Morera's Theorem
one may verify that $Sf$ is an entire holomorphic function on $\mathbb{C}%
^{d}.$ If $f\in L^{p}(\mathbb{R}^{d},\rho)$ with $p>1$ and $Sf=0,$ then $f=0.$
(Reason: if $Sf(i\xi)=0$ for all $\xi\in\mathbb{R}^{d}$ then the Fourier
transform of $f(x)\rho(x)$ is identically zero.)

See Section 4 for a brief discussion of the motivation for the definition of
this transform, and \cite{H4,F} for additional information. Observe from
(\ref{sdef2}) that the Segal-Bargmann transform is very closely related to the
Fourier transform. (In this form, $S$ is essentially the same as the
Fourier-Wiener transform. See \cite{GM} or \cite{H3} for more information on
the connection of the Fourier-Wiener transform to the Segal-Bargmann
transform.) In considering $Sf\left(  x+iy\right)  ,$ one should interpret $x$
as a position variable and $y$ as a frequency variable. In the context of
quantum mechanics, the frequency variable has the interpretation of momentum.

The transform given here is essentially the finite-dimensional version of the
transform described in \cite{S} (see also \cite{BSZ}), and differs from the
transform of \cite{B1} by the ground state transformation. (See Section 4.)
The use of the Gaussian measure instead of the Lebesgue measure of \cite{B1}
is natural in light of the importance of the scale of $L^{p}$ spaces relative
to a Gaussian measure, in connection with hypercontractivity and logarithmic
Sobolev inequalities. (See Section 2.3 and Section 4.) Note that when working
with the Gaussian measure $\rho\left(  x\right)  \,dx,$ the value of $p$ has a
dramatic effect on the allowed growth at infinity of a function in $L^{p}$: a
function in $L^{p}(\mathbb{R}^{d},\rho)$ can grow roughly like $e^{x^{2}/2p}.$

In the case $p=2,$ the image of $L^{p}(\mathbb{R}^{d},\rho)$ under $S$ is
described by the following result \cite{B1,S}.

\begin{theorem}
\label{sb.thm}For any $f\in L^{2}(\mathbb{R}^{d},\rho)$, $Sf$ is a holomorphic
function on $\mathbb{C}^{d}$ and
\[
\int_{\mathbb{R}^{d}}\left|  f\left(  x\right)  \right|  ^{2}\rho\left(
x\right)  \,dx=\pi^{-d}\int_{\mathbb{C}^{d}}\left|  Sf\left(  z\right)
\right|  ^{2}e^{-\left|  z\right|  ^{2}}\,dz.
\]
Conversely, if $F$ is a holomorphic function on $\mathbb{C}^{d}$ for which
\[
\pi^{-d}\int_{\mathbb{C}^{d}}\left|  F\left(  z\right)  \right|
^{2}e^{-\left|  z\right|  ^{2}}\,dz<\infty,
\]
then there is a unique $f\in L^{2}(\mathbb{R}^{d},\rho)$ such that $F=Sf.$
\end{theorem}

The main objective of this paper is to give as precise as possible a
description of the image of $L^{p}(\mathbb{R}^{d},\rho)$ for $p\neq2.$ I know
of no integrability condition that exactly characterizes the image for
$p\neq2.$ Instead we will obtain necessary conditions on the image and
slightly stronger sufficient conditions.

If $F$ is in the image of $L^{p}(\mathbb{R}^{d},\rho)$ then H\"{o}lder's
inequality will tell us that
\begin{equation}
\left|  F(x+iy)\right|  \leq C\,e^{y^{2}/2}e^{x^{2}/2(p-1)}.\label{first.bnd}%
\end{equation}
In the other direction we will show that if $F$ is holomorphic and if
\begin{equation}
\int_{\mathbb{C}^{d}}\left|  F(x+iy)\right|  e^{-y^{2}/2}e^{-x^{2}%
/2(p-1)}\,dx\,dy<\infty\label{first.int}%
\end{equation}
then $F$ is the image of a unique $f$ in $L^{p}(\mathbb{R}^{d},\rho).$ We will
give sharper results for certain ranges of $p.$ Note that the growth condition
in the $y$-direction is the same for all $p,$ but the growth condition in the
$x$-direction is very different for different values of $p.$ These results are
described in Section 2.1.

The sufficient condition (\ref{first.int}) is only polynomially stronger than
the necessary condition in (\ref{first.bnd}). That is, if $F$ satisfies
polynomially better bounds than (\ref{first.bnd}), say
\[
\left|  F(x+iy)\right|  \leq C\,e^{y^{2}/2}e^{x^{2}/2(p-1)}\frac{1}{\left(
1+\left|  x\right|  \right)  ^{d+\varepsilon}\left(  1+\left|  y\right|
\right)  ^{d+\varepsilon}}%
\]
then (\ref{first.int}) will hold and $F$ will be in the image of
$L^{p}(\mathbb{R}^{d},\rho).$ As a consequence of the polynomial closeness of
the two conditions, we will obtain a single necessary-and-sufficient condition
on the image of the ``$L^{p}$ Schwartz space,'' that is, the space of
functions $f\in L^{p}(\mathbb{R}^{d},\rho)$ for which $x^{\alpha}\left(
\partial/\partial x\right)  ^{\beta}f\left(  x\right)  $ is in $L^{p}%
(\mathbb{R}^{d},\rho)$ for all multi-indices $\alpha$ and $\beta.$ This result
generalizes a result of Bargmann \cite{B2} for the case $p=2,$ and is
described in Section 2.2.

Finally, I address the matter of obtaining dimension-independent conditions.
Nelson's hypercontractivity theorem gives us dimension-independent necessary
conditions (for $1<p\leq2$) and sufficient conditions (for $2\leq p<\infty$)
on the image of $L^{p}(\mathbb{R}^{d},\rho).$ These conditions, suitably
interpreted, remain true in the infinite-dimensional ($d=\infty$) case. These
results are described in Section 2.3.

The results in Sections 2.1 and 2.2 are obtained using a one-parameter family
of inversion formulas for $S.$ Since $Sf$ is holomorphic, it is possible to
have many different inversion formulas, each of which expresses $f$ as a
different integral involving the values of $Sf.$ (Just as in Cauchy's formula,
many different integrals involving a holomorphic function $F(z)$ yield the
value $F\left(  z_{0}\right)  .$) For each value of $p$ we choose the
inversion formula that makes best use of growth bounds like $Ce^{y^{2}%
/2}e^{x^{2}/2(p-1)}.$ The ``standard'' inversion formula (i.e. the one given
in \cite{B1,S}) will not give sharp estimates on $f,$ except in the case $p=2.$

The family of inversion formulas is obtained as follows. The space
$L^{p}(\mathbb{R}^{d},e^{-x^{2}/2}\,dx)$ is roughly the same as the space
$L^{2}(\mathbb{R}^{d},e^{-x^{2}/p}\,dx),$ since in either case we can have
growth roughly like $e^{x^{2}/2p}.$ Meanwhile, the results of \cite{H3,DH}
show that $S$ maps $L^{2}(\mathbb{R}^{d},e^{-x^{2}/p}\,dx)$ isometrically onto
the space of holomorphic functions $F$ for which
\[
c\int_{\mathbb{C}^{d}}\left|  F(x+iy)\right|  ^{2}e^{-y^{2}}e^{-x^{2}%
/(p-1)}\,dx\,dy<\infty,
\]
where $c$ is a normalization constant. For each $p$ we get an inversion
formula by saying that the inverse of $S$ is its adjoint, where the adjoint is
computed using on the domain side the inner product in $L^{2}(\mathbb{R}%
^{d},e^{-x^{2}/p}\,dx)$ and on the range side the inner product in
$L^{2}(\mathbb{C}^{d},c\,e^{-y^{2}}e^{-x^{2}/(p-1)}\,dx\,dy).$

It is a pleasure to thank Steve Sontz for valuable scientific discussions and
for making several corrections to the manuscript. This paper was motivated by
the paper \cite{So} of Sontz. Let $\mu$ be the Gaussian measure on
$\mathbb{C}^{d}$ in Proposition \ref{sb.thm}. The paper \cite{So} shows, among
other things, that $S$ maps $L^{p}(\mathbb{R}^{d},\rho)$ into $L^{q}%
(\mathbb{C}^{d},\mu)$ whenever $1\leq q<2$ and $p>1+q/2,$ and that $S$ does
not map $L^{p}(\mathbb{R}^{d},\rho)$ into $L^{q}(\mathbb{C}^{d},\mu)$ if $q>2$
or $p<1+q/2.$ The present paper is an attempt to obtain a more precise
characterization of the image of $L^{p}(\mathbb{R}^{d},\rho).$

\section{Statement of results}

\subsection{Bounds on the transform of $L^{p}$ functions}

\begin{theorem}
\label{holder.thm}Fix $p$ with $1<p<\infty.$ Then for all $f\in L^{p}%
(\mathbb{R}^{d},\rho)$ we have
\[
\left|  Sf\left(  x+iy\right)  \right|  \leq\left\|  f\right\|  _{L^{p}%
(\mathbb{R}^{d},\rho)}e^{y^{2}/2}e^{x^{2}/2(p-1)}.
\]
and
\begin{equation}
\lim_{\left|  x\right|  ^{2}+\left|  y\right|  ^{2}\rightarrow\infty}\left|
Sf\left(  x+iy\right)  \right|  e^{-y^{2}/2}e^{-x^{2}/2(p-1)}%
=0.\label{zero.lim}%
\end{equation}
\end{theorem}

The proof is a simple application of H\"{o}lder's inequality. Note that the
$y$-dependence of the bounds is the same for all $p.$ Roughly speaking, the
rate of growth of $Sf(x+iy)$ in the $x$-direction reflects the rate of growth
of $f\left(  x\right)  ,$ and the rate of growth of $Sf(x+iy)$ in the $y
$-direction reflects the smoothness properties of $f.$ If $f$ is in
$L^{p}(\mathbb{R}^{d},\rho)$ then the allowed growth of $f$ at infinity
depends strongly on $p,$ but the smoothness properties of $f$ depend
comparatively weakly on $p,$ not enough to show up in the above bounds. (See
Theorem \ref{interpol1.thm} for a stronger result if $p\leq2$.)

Going in the opposite direction we have the following result (see also Theorem
\ref{l1lp.thm}).

\begin{theorem}
\label{l1.thm}Fix $p$ with $1<p<\infty.$ Suppose that $F$ is a holomorphic
function on $\mathbb{C}^{d}$ such that
\[
\int_{\mathbb{C}^{d}}\left|  F\left(  x+iy\right)  \right|  e^{-y^{2}%
/2}e^{-x^{2}/2(p-1)}dx\,dy<\infty.
\]
Then there exists a unique $f\in L^{p}(\mathbb{R}^{d},\rho)$ with $Sf=F.$
\end{theorem}

For certain ranges of $p$ we can improve either on the first result or the second.

\begin{theorem}
\label{interpol1.thm}Fix $p$ with $1<p\leq2$ and let $p^{\prime}$ be the
conjugate exponent to $p.$ Then for all $f$ in $L^{p}(\mathbb{R}^{d},\rho)$ we
have
\[
\int_{\mathbb{C}^{d}}\left|  Sf(x+iy)e^{-y^{2}/2}e^{-x^{2}/2(p-1)}\right|
^{p^{\prime}}dx\,dy<\infty.
\]
\end{theorem}

This result gives some control over how fast the expression in (\ref{zero.lim}%
) tends to zero at infinity. Theorem \ref{interpol1.thm} can be viewed as a
sharpening of Theorem 5.1 of \cite{So}, in that it corresponds to the
borderline case in which Sontz's result just fails to apply.

\begin{theorem}
\label{interpol2.thm}Fix $p$ with $2\leq p<\infty$ and let $p^{\prime}$ be the
conjugate exponent to $p.$ Suppose $F$ is a holomorphic function on
$\mathbb{C}^{d}$ such
\[
\int_{\mathbb{C}^{d}}\left|  F(x+iy)e^{-y^{2}/2}e^{-x^{2}/2(p-1)}\right|
^{p^{\prime}}\,dx\,dy<\infty.
\]
Then there exists a unique $f\in L^{p}(\mathbb{R}^{d},\rho)$ with $Sf=F.$
\end{theorem}

We know that to be in the image of $L^{p}(\mathbb{R}^{d},\rho)$ the function
$F$ must satisfy (\ref{zero.lim}). In the presence of this assumption, the
condition in Theorem \ref{interpol2.thm} is easier to satisfy than that in
Theorem \ref{l1.thm}, since $p^{\prime}>1.$

I give one last sufficient condition that in some cases may be better than
Theorem \ref{l1.thm} or Theorem \ref{interpol2.thm}.

\begin{theorem}
\label{l1lp.thm}Fix $p$ with $1<p<\infty.$ Suppose $F$ is a holomorphic
function on $\mathbb{C}^{d}$ such that
\[
\int_{\mathbb{R}^{d}}\left(  \int_{\mathbb{R}^{d}}\left|  F(x+iy)\right|
e^{-y^{2}/2}e^{-x^{2}/2(p-1)}dy\right)  ^{p}\,dx<\infty.
\]
Then there exists a unique $f$ in $L^{p}(\mathbb{R}^{d},\rho)$ with $Sf=F.$
\end{theorem}

\subsection{Bounds on the transform of the $L^{p}$ Schwartz space}

In this subsection we consider the ``$L^{p}$ Schwartz space,'' that is, the
space of all $f$ in $L^{p}(\mathbb{R}^{d},\rho)$ such that any derivative of
$f$ times any polynomial is again in $L^{p}(\mathbb{R}^{d},\rho).$ The
closeness of the necessary conditions and the sufficient conditions in the
previous subsection allow us to give a single necessary-and-sufficient
pointwise condition on the image of this space.

\begin{theorem}
\label{schwarz.thm}Let $p$ be a number with $1<p<\infty.$ Suppose $F$ is a
holomorphic function on $\mathbb{C}^{d}.$ Then the following two conditions
are equivalent.

\begin{enumerate}
\item  For all $n\in\mathbb{N}$ there exists a constant $C_{n}$ such that
\[
\left|  F\left(  x+iy\right)  \right|  \leq C_{n}\,e^{y^{2}/2}e^{x^{2}%
/2(p-1)}\frac{1}{(1+\left|  x\right|  )^{n}(1+\left|  y\right|  )^{n}}.
\]

\item  There exists $f\in L^{p}(\mathbb{R}^{d},\rho)$ such that $Sf=F$ where
$f$ is smooth and has the property that for all multi-indices $\alpha,\beta$%
\[
x^{\alpha}\left(  \frac{\partial}{\partial x}\right)  ^{\beta}f\left(
x\right)  \in L^{p}(\mathbb{R}^{d},\rho).
\]
\end{enumerate}
\end{theorem}

\textit{Remarks}. 1) The proof will show that if $F$ satisfies Condition 1
then so does $(\partial/\partial z)^{\alpha}F$ for all multi-indices $\alpha.$
See the discussion after the proof of Theorem \ref{schwarz.thm}.

2) The space of functions satisfying Condition 2 of the theorem may be thought
of as a sort of ``Schwartz space'' associated to $L^{p}(\mathbb{R}^{d},\rho).$
In the case $p=2,$ the space of functions satisfying Condition 2 is precisely
the image of the Schwartz space under the ground state transformation (see
Lemma \ref{schwarz.lem}). Thus the $p=2$ case of Theorem \ref{schwarz.thm}
follows from \cite[Thm. 1.7]{B2}.

\subsection{Dimension-independent bounds}

Theorem \ref{holder.thm} has dimension-independent constants and so remains
true, when suitably interpreted, in the infinite-dimensional ($d=\infty$)
case. (See \cite{BSZ}, \cite{GM}, or \cite[Sect. 10]{H4} for information on
the infinite-dimensional form of the Segal-Bargmann transform.) However, the
other results of Section 2.2, and the results of Section 2.3, are completely
dimension-dependent and do not carry over to the infinite-dimensional case.
Nevertheless, hypercontractivity (e.g. \cite{DGS}) does give us some
dimension-independent estimates, as follows.

\begin{theorem}
\label{hyper1.thm}Fix $p$ with $1<p\leq2.$ Then for all $f\in L^{p}%
(\mathbb{R}^{d},\rho)$ we have
\begin{equation}
\int_{\mathbb{C}^{d}}\left|  Sf\left(  \sqrt{p-1}z\right)  \right|  ^{2}%
\frac{e^{-\left|  z\right|  ^{2}}}{\pi^{d}}\,dz\leq\left\|  f\right\|
_{L^{p}(\mathbb{R}^{d},\rho)}^{2}.\label{hyper1}%
\end{equation}
\end{theorem}

\begin{theorem}
\label{hyper2.thm}Fix $p$ with $2\leq p<\infty.$ Suppose $F$ is a holomorphic
function on $\mathbb{C}^{d}$ such that
\begin{equation}
\int_{\mathbb{C}^{d}}\left|  F\left(  \sqrt{p-1}z\right)  \right|  ^{2}%
\frac{e^{-\left|  z\right|  ^{2}}}{\pi^{d}}\,dz=c^{2}<\infty.\label{hyper2}%
\end{equation}
Then there exists a unique $f\in L^{p}(\mathbb{R}^{d},\rho)$ such that $Sf=F,$
and
\[
\left\|  f\right\|  _{L^{p}(\mathbb{R}^{d},\rho)}\leq c.
\]
\end{theorem}

The main virtue of these results is their dimension-independence. If we
compare them for fixed finite $d$ to Theorems \ref{interpol1.thm} and
\ref{interpol2.thm}, we can say that Theorems \ref{hyper1.thm} and
\ref{hyper2.thm} are slightly better concerning the behavior in the
$x$-direction and much worse concerning behavior in the $y$-direction.

Consider, for example, the case $p<2.$ Equation (\ref{hyper1}) tells us, after
making the change of variable $u=\sqrt{p-1}z,$ that for $F$ in the image of
$L^{p}(\mathbb{R}^{d},\rho)$ we have
\[
\int_{F}\left|  F(x+iy)e^{-x^{2}/2(p-1)}\right|  ^{2}\,dx<\infty
\]
for almost every $y,$ whereas from Theorem \ref{interpol1.thm} we can only
conclude the same result with the exponent $2$ replaced by $p^{\prime}>2.$ On
the other hand, a holomorphic function $F$ satisfying
\[
\int_{\mathbb{C}^{d}}\left|  F\left(  \sqrt{p-1}z\right)  \right|  ^{2}%
\frac{e^{-\left|  z\right|  ^{2}}}{\pi^{d}}\,dz<\infty
\]
can grow roughly like $e^{y^{2}/2(p-1)}$ in the $y$-direction, whereas Theorem
\ref{interpol1.thm} allows only growth like $e^{y^{2}/2}.$

Similarly for $p>2,$ the condition for Theorem \ref{hyper2.thm} requires
slightly weaker growth conditions in the $x$-direction than Theorem
\ref{interpol2.thm} (because $2>p^{\prime}$), but much stronger conditions in
the $y$-direction ($e^{y^{2}/2(p-1)}$ versus $e^{y^{2}/2}$).

It should be pointed out that in the infinite-dimensional case, bounds do not
imply integrability. That is, even having bounds like $\left|  F(x+iy)\right|
\leq Ce^{\varepsilon(x^{2}+y^{2)}}$ for some small $\varepsilon$ is
insufficient to guarantee that $F$ is in the Segal-Bargmann space. Thus in the
infinite-dimensional case it is useful to have both pointwise bounds (from
Theorem \ref{holder.thm}) and integrability conditions (from Theorem
\ref{hyper1.thm}) on the image of $L^{p}.$

\section{Proofs}

\subsection{Bounds on the transform of $L^{p}$ functions}

\textit{Proof of Theorem \ref{holder.thm}}. Formula (\ref{sdef2}) expresses
the Segal-Bargmann transform as
\[
Sf\left(  z\right)  =e^{-z^{2}/2}\int_{\mathbb{R}^{d}}e^{z\cdot x}f\left(
x\right)  \rho\left(  x\right)  \,dx.
\]
Thus by H\"{o}lder's Inequality,
\begin{equation}
\left|  Sf\left(  z\right)  \right|  \leq\left|  e^{-z^{2}/2}\right|  \left\|
f\right\|  _{L^{p}(\mathbb{R}^{d},\rho)}\left\|  e^{z\cdot x}\right\|
_{L^{q}(\mathbb{R}^{d},\rho)}\label{holder}%
\end{equation}
where $1/p+1/q=1.$ Now, $\left|  e^{z\cdot x}\right|  =e^{\operatorname{Re}%
z\cdot x}.$ Completing the square gives
\begin{align*}
\int_{\mathbb{R}^{d}}\left|  e^{z\cdot x}\right|  ^{q}\rho\left(  x\right)
\,dx &  =\left(  2\pi\right)  ^{-d/2}\int_{\mathbb{R}^{d}}%
e^{q\operatorname{Re}z\cdot x}e^{-x^{2}/2}\,dx\\
&  =e^{q^{2}(\operatorname{Re}z)^{2}/2}\left(  2\pi\right)  ^{-d/2}%
\int_{\mathbb{R}^{d}}e^{-(x-q\operatorname{Re}z)^{2}/2}\,dx\\
&  =e^{q^{2}(\operatorname{Re}z)^{2}/2}.
\end{align*}
Thus $\left\|  e^{z\cdot x}\right\|  _{L^{q}(\mathbb{R}^{d},\rho
)}=e^{q(\operatorname{Re}z)^{2}/2}.$ Then simplifying the bounds in
(\ref{holder}) gives the estimate in Theorem \ref{holder.thm}.

Now if $f\in L^{p}(\mathbb{R}^{d},\rho)$ is a polynomial, then using
(\ref{sdef1}) we can see that $Sf$ is a polynomial, so the expression
\begin{equation}
\left|  Sf\left(  x+iy\right)  \right|  e^{-y^{2}/2}e^{-x^{2}/2(p-1)}%
\label{lim.quant}%
\end{equation}
will certainly tend to zero at infinity. For any $f$ in $L^{p}(\mathbb{R}%
^{d},\rho)$ we can approximate $f$ in the $L^{p}$ norm by a sequence $f_{n}$
of polynomials. By the bounds just established, the expression
(\ref{lim.quant}) for $f_{n}$ will converge uniformly to the same expression
for $f.$ So in the limit (\ref{lim.quant}) still tends to zero at infinity,
establishing (\ref{zero.lim}). $\square$

\textit{Proof of Theorem \ref{l1.thm}}. The uniqueness of $f$ follows from the
injectivity of $S$ on $L^{p}(\mathbb{R}^{d},\rho),$ discussed in the
introduction. For the existence we will make use of an inversion formula
obtained from the following result.

\begin{theorem}
\label{st.thm}For all $p>1,$ the Segal-Bargmann transform $S$ is a unitary map
of $L^{2}(\mathbb{R}^{d},\rho_{p/2})$ onto $\mathcal{H}L^{2}(\mathbb{C}%
^{d},\mu_{p/2}),$ where
\[
\rho_{p/2}(x)=\left(  \pi p\right)  ^{-d/2}e^{-x^{2}/p}%
\]
and where
\[
\mu_{p/2}\left(  x+iy\right)  =\left(  \pi(p-1)\right)  ^{-d/2}\pi
^{-d/2}e^{-x^{2}/(p-1)}e^{-y^{2}}.
\]
Here $\mathcal{H}L^{2}(\mathbb{C}^{d},\mu_{p/2})$ denotes the space of
holomorphic functions on $\mathbb{C}^{d}$ that are square-integrable with
respect to $\mu_{p/2}\left(  x+iy\right)  \,dx\,dy.$
\end{theorem}

This is Theorem 3.2 of \cite{DH} (with $s=p/2$ and $t=1$) and a special case
of Theorem 1.2 of \cite{H3}. (See also \cite{Sen}.) Note that if $p=2,$ then
$\rho_{p/2}$ is just the density $\rho,$ and similarly for $\mu_{p/2}.$

As discussed in the introduction, $L^{p}(\mathbb{R}^{d},\rho)$ is roughly the
same as $L^{2}(\mathbb{R}^{d},\rho_{p/2}),$ so we expect the image of
$L^{p}(\mathbb{R}^{d},\rho)$ to be roughly the same as the image of
$L^{2}(\mathbb{R}^{d},\rho_{p/2}),$ namely, $\mathcal{H}L^{2}(\mathbb{C}%
^{d},\mu_{p/2}).$ For each $p$ there is an inversion formula for $S$ that maps
$\mathcal{H}L^{2}(\mathbb{C}^{d},\mu_{p/2})$ isometrically onto $L^{2}%
(\mathbb{R}^{d},\rho_{p/2})$ and is zero on the orthogonal complement of
$\mathcal{H}L^{2}(\mathbb{C}^{d},\mu_{p/2})$ in $L^{2}(\mathbb{C}^{d}%
,\mu_{p/2}).$ This inversion formula gives good estimates on $S^{-1}F$ when
$F$ has growth like $e^{y^{2}/2}e^{x^{2}/2(p-1)}.$

Specifically, we write $Sf$ in the form
\[
Sf\left(  z\right)  =\left(  2\pi\right)  ^{-d/2}\int\frac{e^{-(z-x)^{2}/2}%
}{\rho_{p/2}\left(  x\right)  }f\left(  x\right)  \rho_{p/2}\left(  x\right)
\,dx.
\]
We now take the adjoint of $S,$ viewed as an operator from the Hilbert space
$L^{2}(\mathbb{R}^{d},\rho_{p/2})$ into the Hilbert space $L^{2}%
(\mathbb{C}^{d},\mu_{p/2}),$ whose image is the holomorphic subspace of
$L^{2}(\mathbb{C}^{d},\mu_{p/2}).$ Since $S$ is isometric, its adjoint is a
one-sided inverse to $S.$ On the other hand, since we have expressed $S$ as an
integral operator, its adjoint may be computed in the usual way by taking the
complex conjugate of the integral kernel and reversing the roles of the
variables. So we obtain on operator $S^{\ast,p}:L^{2}(\mathbb{C}^{d},\mu
_{p/2})\rightarrow L^{2}(\mathbb{R}^{d},\rho_{p/2})$ given by
\begin{align}
S^{\ast,p}F(x) &  =\left(  2\pi\right)  ^{-d/2}\int_{\mathbb{C}^{d}}%
\frac{e^{-(\bar{z}-x)^{2}/2}}{\rho_{p/2}\left(  x\right)  }F\left(  z\right)
\mu_{p/2}\left(  z\right)  \,dz\nonumber\\
&  =ce^{x^{2}/p}\int_{\mathbb{R}^{d}}\int_{\mathbb{R}^{d}}e^{-\left(
\tilde{x}-i\tilde{y}-x\right)  ^{2}/2}F\left(  \tilde{x}+i\tilde{y}\right)
e^{-\tilde{x}^{2}/(p-1)}e^{-\tilde{y}^{2}}d\tilde{x}\,d\tilde{y}%
.\label{inverse}%
\end{align}
Here $c$ is a constant which depends on $p$ and the dimension.

More precisely, the formula (\ref{inverse}) for $S^{\ast,p}$ makes sense
whenever all the integrals are convergent, for example if $F$ has compact
support. For general $F$ in $L^{2}(\mathbb{C}^{d},\mu_{p/2})$ we compute
$S^{\ast,p}$ by integrating over a compact set and then taking a limit in
$L^{2}(\mathbb{R}^{d},\rho_{p/2}).$ If $F$ is in the holomorphic subspace of
$L^{2}(\mathbb{C}^{d},\mu_{p/2})$ then since $S$ is isometric $SS^{\ast
,p}F=F,$ so that $S^{\ast,p}$ is computing the inverse of $S.$

We now re-write the integral (\ref{inverse}) so that it is expressed in terms
of the Gaussian measure in Theorem \ref{l1.thm}. So
\begin{align*}
S^{\ast,p}F(x)  & =c\int_{\mathbb{R}^{d}}\int_{\mathbb{R}^{d}}e^{x^{2}%
/p}e^{-\left(  \tilde{x}-i\tilde{y}-x\right)  ^{2}/2}e^{-\tilde{x}^{2}%
/2(p-1)}e^{-\tilde{y}^{2}/2}\cdot\\
& \cdot\left[  F\left(  \tilde{x}+i\tilde{y}\right)  e^{-\tilde{x}^{2}%
/2(p-1)}e^{-\tilde{y}^{2}/2}\right]  d\tilde{x}\,d\tilde{y}.
\end{align*}
A straightforward Gaussian integral shows that
\[
\left\|  e^{x^{2}/p}e^{-\left(  \tilde{x}-i\tilde{y}-x\right)  ^{2}%
/2}e^{-\tilde{x}^{2}/2(p-1)}e^{-\tilde{y}^{2}/2}\right\|  _{L^{p}%
(\mathbb{R}^{d},\rho)}=const.
\]
independent of $\tilde{x}$ and $\tilde{y},$ where the norm is computed with
respect to the $x$-variable. So if $F$ is in $L^{2}(\mathbb{C}^{d},\mu_{p/2})$
and such that (\ref{inverse}) converges absolutely, we obtain by putting the
$L^{p}(\mathbb{R}^{d},\rho)$ norm inside the integral that
\begin{equation}
\left\|  S^{\ast,p}F\right\|  _{L^{p}(\mathbb{R}^{d},\rho)}\leq c\,\int
_{\mathbb{C}^{d}}\left|  F\left(  \tilde{x}+i\tilde{y}\right)  \right|
e^{-\tilde{x}^{2}/2(p-1)}e^{-\tilde{y}^{2}/2}\,d\tilde{x}\,d\tilde
{y}.\label{int.l1}%
\end{equation}

This estimate shows that $S^{\ast,p}$ extends to a continuous map of
\begin{equation}
L^{1}(\mathbb{C}^{d},e^{-\tilde{x}^{2}/2(p-1)}e^{-\tilde{y}^{2}/2}\,d\tilde
{x}\,d\tilde{y})\label{l1.space}%
\end{equation}
into $L^{p}(\mathbb{R}^{d},\rho).$ I claim that if $F$ is holomorphic and in
this $L^{1}$ space, then $SS^{\ast,p}F=F.$ To see this, approximate $F$ by
$F_{n}(z)=F(\lambda_{n}z),$ with $\lambda_{n}$ tending to one from below. Then
$F_{n}$ will converge to $F$ in the $L^{1}$ space (\ref{l1.space}), and using
elementary pointwise bounds \cite[Sect. 2]{H4} we can establish that each
$F_{n}$ is in $\mathcal{H}L^{2}(\mathbb{C}^{d},\mu_{p/2}),$ so that
$SS^{\ast,p}F_{n}=F_{n.}$ Then by (\ref{int.l1}) $S^{\ast,p}F_{n}$ will
converge to $S^{\ast,p}F$ in $L^{p}(\mathbb{R}^{d},\rho),$ so that
$SS^{\ast,p}F_{n}=F_{n}$ will converge uniformly on compact sets to
$SS^{\ast,p}F.$ Since $F_{n}$ also converges in the $L^{1}$ space to $F,$ the
pointwise limit and the $L^{1}$ limit must coincide, so $SS^{\ast,p}F=F. $

Thus for any holomorphic function $F$ in the $L^{1}$ space (\ref{l1.space}),
there exists a function $f$ in $L^{p}(\mathbb{R}^{d},\rho)$ (namely,
$f=S^{\ast,p}F$) for which $Sf=F.$ This establishes Theorem \ref{l1.thm}.
$\square$

\textit{Proof of Theorem \ref{interpol1.thm}}. We have stated (Theorem
\ref{st.thm}) that for each $p\in(1,\infty)$, $S$ is a unitary map of
$L^{2}(\mathbb{R}^{d},\rho_{p/2})$ onto $\mathcal{H}L^{2}(\mathbb{C}^{d}%
,\mu_{p/2}).$ We can then construct an isometric Lebesgue measure map
\[
S_{p}:L^{2}(\mathbb{R}^{d},dx)\rightarrow L^{2}(\mathbb{C}^{d},dx\,dy)
\]
by making a change of measure on both sides. That is, $S_{p}$ is defined to
by
\[
S_{p}f(z)=\mu_{p/2}(z)^{1/2}S(\rho_{p/2}(x)^{-1/2}f(x)).
\]

Computing explicitly we obtain, after some algebra,
\begin{align}
S_{p}f\left(  \tilde{x}+i\tilde{y}\right)   & =c\,e^{-\tilde{y}^{2}%
/2}e^{-\tilde{x}^{2}/2(p-1)}\int_{\mathbb{R}^{d}}e^{-z^{2}/2}e^{z\cdot
x}\left(  e^{x^{2}/2p}f(x)\right)  e^{-x^{2}/2}\,dx\nonumber\\
& =c\int_{\mathbb{R}^{d}}e^{-i\tilde{x}\cdot\tilde{y}}e^{ix\cdot\tilde{y}}%
\exp\left\{  -\frac{p-1}{2p}\left[  x-\frac{p}{p-1}\tilde{x}\right]
^{2}\right\}  f\left(  x\right)  \,dx,\label{sp.form}%
\end{align}
where $z=\tilde{x}+i\tilde{y}$ and $c$ is a constant depending on $p$ and the
dimension. Clearly $S_{p}$ also makes sense on $L^{1}(\mathbb{R}^{d},dx)$ and
defines a bounded linear map of $L^{1}(\mathbb{R}^{d},dx)$ into $L^{\infty
}(\mathbb{C}^{d}).$ Therefore by interpolation (e.g. \cite[Thm. V.1.3]{SW}),
$S_{p}$ is a bounded map of $L^{q}(\mathbb{R}^{d},dx)$ into $L^{q^{\prime}%
}(\mathbb{C}^{d},dx\,dy)$ for all $q$ with $1\leq q\leq2.$

If $1<p\leq2,$ then there is nothing to prevent us from taking $q=p,$ so that
$S_{p}$ is bounded from $L^{p}(\mathbb{R}^{d},dx)$ into $L^{p^{\prime}%
}(\mathbb{C}^{d},dx\,dy).$ Let us then make another change of measure on both
sides. On the $\mathbb{R}^{d}$ side we change from $L^{p}(\mathbb{R}^{d},dx)$
to $L^{p}(\mathbb{R}^{d},\rho),$ by multiplying by $\rho(x)^{1/p}.$ This is,
up to a constant, the reverse of the change of measure in the construction of
$S_{p},$ since $\rho_{p/2}(x)^{1/2}=c_{1}\rho(x)^{1/p}=c_{2}e^{-x^{2}/2p}.$ On
the $\mathbb{C}^{d}$ side we want to change from $L^{p^{\prime}}%
(\mathbb{C}^{d},dx\,dy)$ to $L^{p^{\prime}}(\mathbb{C}^{d},\mu_{p/2}%
^{p^{\prime}})$ where
\begin{align*}
\mu_{p/2}^{p^{\prime}}(x+iy)  & =\left(  \mu_{p/2}(x+iy)\right)  ^{p^{\prime}%
}\\
& =c\,e^{-p^{\prime}y^{2}/2}e^{-p^{\prime}x^{2}/2(p-1)}dx\,dy.
\end{align*}
After all, $L^{p^{\prime}}(\mathbb{C}^{d},\mu_{p/2}^{p^{\prime}})$ is the
space that appears (implicitly) in Theorem \ref{interpol1.thm}. This change of
measure is accomplished by dividing by $e^{-y^{2}/2}e^{-x^{2}/2(p-1)},$ which
again is up to a constant the reverse of the change of measure in the
construction of $S_{p}.$ The result is that if we make these changes of
measure we simply get back $S$ again, up to an irrelevant constant.

We conclude then that $S$ is a bounded map of $L^{p}(\mathbb{R}^{d},\rho)$
into $L^{p^{\prime}}(\mathbb{C}^{d},\mu_{p/2}^{p^{\prime}}),$ which is the
content of Theorem \ref{interpol1.thm}. $\square$

\textit{Proof of Theorem \ref{interpol2.thm}}. Now consider the bounded map
$S^{\ast,p}:L^{2}(\mathbb{C}^{d},\mu_{p/2})\rightarrow L^{2}(\mathbb{R}%
^{d},\rho_{p/2})$ given by (\ref{inverse}). Recall that if $F$ is in the
holomorphic subspace of $L^{2}(\mathbb{C}^{d},\mu_{p/2})$ then $S^{\ast
,p}F=S^{-1}F.$ If we make a change of measure to Lebesgue measure on both
sides, we obtain simply the adjoint $S_{p}^{\ast}$ of the map $S_{p}$ given by
(\ref{sp.form}). So $S_{p}^{\ast}$ is given by
\[
S_{p}^{\ast}F(x)=c\int_{\mathbb{R}^{d}}e^{i\tilde{x}\cdot\tilde{y}}%
e^{-ix\cdot\tilde{y}}\exp\left\{  -\frac{p-1}{2p}\left[  x-\frac{p}{p-1}%
\tilde{x}\right]  ^{2}\right\}  F(\tilde{x}+i\tilde{y})\,d\tilde{x}%
\,d\tilde{y}%
\]
(on a suitable dense subspace, and then extended by continuity). Clearly
$S_{p}^{\ast}$ makes sense on $L^{1}(\mathbb{C}^{d},dx\,dy)$ and maps
$L^{1}(\mathbb{C}^{d},dx\,dy)$ continuously into $L^{\infty}(\mathbb{R}^{d}).$
So by interpolation again, $S_{p}^{\ast}$ maps $L^{q}(\mathbb{C}^{d},dx\,dy)$
boundedly into $L^{q^{\prime}}(\mathbb{R}^{d},dx)$ for all $q$ with $1\leq
q\leq2.$ If $p\geq2,$ we may take $q=p^{\prime}$ so that $S_{p}^{\ast}$ maps
$L^{p^{\prime}}(\mathbb{C}^{d},dx\,dy)$ boundedly into $L^{p}(\mathbb{R}^{d},dx).$

We now change measure again, as in the proof of Theorem \ref{interpol1.thm},
to $L^{p^{\prime}}(\mathbb{C}^{d},\mu_{p/2}^{p^{\prime}})$ on the complex side
and to $L^{p}(\mathbb{R}^{d},\rho)$ on the real side. Again these changes of
measure simply undo (up to a constant) the changes made to get $S_{p}^{\ast}$
from $S^{\ast,p}.$ So we conclude that $S^{\ast,p}$ maps $L^{p^{\prime}%
}(\mathbb{C}^{d},\mu_{p/2}^{p^{\prime}})$ boundedly into $L^{p}(\mathbb{R}%
^{d},\rho).$ Arguing as in the last part of the proof of Theorem \ref{l1.thm},
we see that if $F$ is in the holomorphic subspace of $L^{p^{\prime}%
}(\mathbb{C}^{d},\mu_{p/2}^{p^{\prime}}),$ then $SS^{\ast,p}F=F.$ Thus for all
holomorphic $F$ in $L^{p^{\prime}}(\mathbb{C}^{d},\mu_{p/2}^{p^{\prime}})$
($p\geq2$), there is $f$ in $L^{p}(\mathbb{R}^{d},\rho)$ with $Sf=F,$ namely,
$f=S^{\ast,p}F.$ This is the content of Theorem \ref{interpol2.thm}. $\square$

\textit{Proof of Theorem \ref{l1lp.thm}}. We use the inversion formula
(\ref{inverse}). Putting absolute values inside the integral and simplifying
gives
\begin{align*}
& \left|  S^{\ast,p}F(x)\right|  \\
& \leq c\,e^{x^{2}/2p}\int_{\mathbb{R}^{d}}e^{-\frac{p}{2(p-1)}\left[
\tilde{x}-\frac{p-1}{p}x\right]  ^{2}}\int_{\mathbb{R}^{d}}\left|  F(\tilde
{x}+i\tilde{y})e^{-\tilde{y}^{2}/2}e^{-\tilde{x}^{2}/2(p-1)}\right|
\,d\tilde{y}\,d\tilde{x}.
\end{align*}
Suppose now that the inner integral
\begin{equation}
\int_{\mathbb{R}^{d}}\left|  F(\tilde{x}+i\tilde{y})e^{-\tilde{y}^{2}%
/2}e^{-\tilde{x}^{2}/2(p-1)}\right|  \,d\tilde{y}\label{lp.inner}%
\end{equation}
is in $L^{p}(\mathbb{R}^{d},dx).$ Then the outer integral is simply computing
the convolution of an $L^{p}$ function with a Gaussian, so by the standard
inequality
\[
\left\|  f\ast g\right\|  _{L^{p}(\mathbb{R}^{d},dx)}\leq\left\|  f\right\|
_{L^{p}(\mathbb{R}^{d},dx)}\left\|  g\right\|  _{L^{1}(\mathbb{R}^{d},dx)}%
\]
we conclude that outer integral is in $L^{p}(\mathbb{R}^{d},dx)$ as well. This
implies that $e^{-x^{2}/2p}S^{\ast,p}F(x)$ is in $L^{p}(\mathbb{R}^{d},dx),$
which means that $S^{\ast,p}F$ is in $L^{p}(\mathbb{R}^{d},\rho).$ So, if
(\ref{lp.inner}) is in $L^{p}(\mathbb{R}^{d},dx),$ then $S^{\ast,p}F$ is in
$L^{p}(\mathbb{R}^{d},\rho).$ As in the previous proofs, we can show that if
in addition $F$ is holomorphic, then $SS^{\ast,p}F=f,$ so that $F$ is the
transform of something in $L^{p}(\mathbb{R}^{d},\rho).$ $\square$

\subsection{Bounds on the transform of the $L^{p}$ Schwartz space}

We begin with the following elementary result.

\begin{lemma}
\label{schwarz.lem}If $f$ is a smooth function on $\mathbb{R}^{d}$ such that
$x^{\alpha}(\partial/\partial x)^{\beta}f\in L^{p}(\mathbb{R}^{d},\rho)$ for
all $\alpha$ and $\beta,$ then for all $\alpha$ and $\beta$ there exists a
constant $c_{\alpha,\beta}$ such that
\[
x^{\alpha}\left(  \frac{\partial}{\partial x}\right)  ^{\beta}f\left(
x\right)  \leq c_{\alpha,\beta}e^{x^{2}/2p}.
\]
\end{lemma}

\textit{Proof}. If $f\in L^{p}(\mathbb{R}^{d},\rho)$ then $f\left(  x\right)
e^{-x^{2}/2p}\in L^{p}(\mathbb{R}^{d},dx).$ Note that all of the derivatives
of $e^{-x^{2}/2p}$ are polynomials times $e^{-x^{2}/2p}.$ Thus if $x^{\alpha
}(\partial/\partial x)^{\beta}f$ is in $L^{p}(\mathbb{R}^{d},\rho)$ for all
$\alpha$ and $\beta,$ then $x^{\alpha}(\partial/\partial x)^{\beta
}(f(x)e^{-x^{2}/2p})$ is in $L^{p}(\mathbb{R}^{d},dx)$ for all $\alpha$ and
$\beta.$ Thus by standard Sobolev embedding theorems, $f(x)e^{-x^{2}/2p}$ is a
Schwartz function. Using again that the derivatives of $e^{-x^{2}/2p}$ are
polynomials times $e^{-x^{2}/2p},$ we see that $x^{\alpha}[(\partial/\partial
x)^{\beta}f(x)]e^{-x^{2}/2p}$ is bounded for all $\alpha$ and $\beta,$ and the
Lemma follows. $\square$

\textit{Proof of Theorem \ref{schwarz.thm}}. We first prove that Condition 2
in the theorem implies Condition 1. So assume that $F$ is a holomorphic
function that is of the form $F=Sf,$ where $f$ has the property given in
Condition 2. Since $p>1,$ the function $\ e^{-\left(  z-x\right)  ^{2}%
/2}e^{x^{2}/2p}$ tends to zero rapidly as $x$ tends to infinity (for each
fixed $z\in\mathbb{C}^{d}$). Thus Condition 2 and Lemma \ref{schwarz.lem}
permit us to integrate by parts with no boundary terms, giving
\begin{align*}
S\frac{\partial f}{\partial x_{k}}\left(  z\right)   &  =\left(  2\pi\right)
^{-d/2}\int_{\mathbb{R}^{d}}e^{-\left(  z-x\right)  ^{2}/2}\frac{\partial
f}{\partial x_{k}}\,dx\\
&  =-\left(  2\pi\right)  ^{-d/2}\int_{\mathbb{R}^{d}}(z_{k}-x_{k})e^{-\left(
z-x\right)  ^{2}/2}f\left(  x\right)  \,dx.
\end{align*}
So
\[
S\frac{\partial f}{\partial x_{k}}=-z_{k}Sf+S\left[  x_{k}f\left(  x\right)
\right]
\]
or
\begin{equation}
S\left(  x_{k}-\frac{\partial}{\partial x_{k}}\right)  f=z_{k}Sf=z_{k}%
F.\label{create}%
\end{equation}

Now the assumptions on $f$ tell us that $(x_{1}-\partial/\partial
x_{1})^{n_{1}}\cdots(x_{d}-\partial/\partial x_{d})^{n_{d}}f$ is in
$L^{p}(\mathbb{R}^{d},\rho)$ for all positive integers $n_{1},\cdots,n_{d}.$
So applying (\ref{create}) repeatedly we see that $z_{1}^{n_{1}}\cdots
z_{d}^{n_{d}}F\left(  z\right)  $ is the transform of a function in
$L^{p}(\mathbb{R}^{d},\rho).$ So both $F$ itself and $z_{1}^{n_{1}}\cdots
z_{d}^{n_{d}}F\left(  z\right)  $ satisfy the bounds given in Theorem
\ref{holder.thm}, from which the bounds given in Condition 1 of Theorem
\ref{schwarz.thm} easily follow.

We now prove that Condition 1 implies Condition 2. Assume that $F$ is
holomorphic and satisfies the bounds of Condition 1. Then by direct
calculation $F$ is square-integrable with respect the measure $\mu
_{p/2}\left(  z\right)  \,dz$ and therefore by Theorem \ref{st.thm} there
exists a function $f$ with $Sf=F.$ We may calculate $f$ by the inversion
formula (\ref{inverse}). Putting absolute values inside the integral and
simplifying gives (as noted previously)
\begin{align}
& \left|  S^{-1}F(x)\right|  \nonumber\\
& \leq c\,e^{x^{2}/2p}\int_{\mathbb{R}^{d}}e^{-\frac{p}{2(p-1)}\left[
\tilde{x}-\frac{p-1}{p}x\right]  ^{2}}\int_{\mathbb{R}^{d}}\left|  F(\tilde
{x}+i\tilde{y})e^{-\tilde{y}^{2}/2}e^{-\tilde{x}^{2}/2(p-1)}\right|
\,d\tilde{y}\,d\tilde{x}.\label{direct.bnd}%
\end{align}
Substituting in the bounds then gives (for $n>d$)
\begin{align*}
\left|  S^{-1}F(x)\right|    & \leq C_{n}\,e^{x^{2}/2p}\int_{\mathbb{C}^{d}%
}e^{-\frac{p}{2(p-1)}\left[  \tilde{x}-\frac{p-1}{p}x\right]  ^{2}}\frac
{1}{(1+\left|  \tilde{x}\right|  )^{n}(1+\left|  \tilde{y}\right|  )^{n}%
}\,d\tilde{x}\,d\tilde{y}\\
& \leq D_{n}\,e^{x^{2}/2p}\int_{\mathbb{C}^{d}}e^{-\frac{p}{2(p-1)}\left[
\tilde{x}-\frac{p-1}{p}x\right]  ^{2}}\frac{1}{(1+\left|  \tilde{x}\right|
)^{n}}\,d\tilde{x}%
\end{align*}
In particular the bounds guarantee that the integral is absolutely convergent.

Now, it is not difficult to see that the integral
\[
\int_{\mathbb{C}^{d}}e^{-\frac{p}{2(p-1)}\left[  \tilde{x}-\frac{p-1}%
{p}x\right]  ^{2}}\frac{1}{(1+\left|  \tilde{x}\right|  )^{n}}\,d\tilde
{x}\,d\tilde{y}%
\]
decays like a constant times $\left|  x\right|  ^{-n}$ as $x$ tends to
infinity. (The main contribution to the integral comes from the region near
where $\tilde{x}=(p-1)x/p,$ and at that point the integrand is of order
$\left|  x\right|  ^{-n}.$) Thus $S^{-1}F(x)$ is bounded by $c_{n}e^{x^{2}%
/2p}(1+\left|  x\right|  )^{-n}$ for all $n.$ This shows that any polynomial
times $f$ is in $L^{p}(\mathbb{R}^{d},\rho).$

We need then to consider the derivatives of $f.$ Using again the inversion
formula (\ref{inverse}) we may write
\[
f=\lim_{R\rightarrow\infty}f_{R}%
\]
where
\[
f_{R}(x)=ce^{x^{2}/p}\int_{\left|  \tilde{x}_{l}\right|  \leq R}\int_{\left|
\tilde{y}_{l}\right|  \leq R}e^{-\left(  \bar{z}-x\right)  ^{2}/2}F\left(
z\right)  e^{-\tilde{x}^{2}/(p-1)}e^{-\tilde{y}^{2}}d\tilde{x}\,d\tilde{y}.
\]
Here $z=\tilde{x}+i\tilde{y}$ and $\left|  \tilde{x}_{l}\right|  \leq R$ means
that we integrate over the set of points $\tilde{x}=\left(  \tilde{x}%
_{1},\cdots,\tilde{x}_{d}\right)  $ in $\mathbb{R}^{d}$ where each coordinate
is at most $R$ in absolute value. Differentiating this gives
\begin{align*}
\frac{\partial f_{R}}{\partial x_{k}}  &  =ce^{x^{2}/p}\int_{\left|  \tilde
{x}_{l}\right|  \leq R}\int_{\left|  \tilde{y}_{l}\right|  \leq R}\left(
\frac{\partial}{\partial x_{k}}e^{-\left(  \bar{z}-x\right)  ^{2}/2}\right)
F\left(  z\right)  e^{-\tilde{x}^{2}/(p-1)}e^{-\tilde{y}^{2}}d\tilde
{x}\,d\tilde{y}\\
&  +\frac{2x_{k}}{p}f_{R}\left(  x\right)  .
\end{align*}

Note that $(\partial/\partial x_{k})\exp\left[  -(\bar{z}-x)^{2}/2\right]
=-(\partial/\partial\bar{z}_{k})\exp\left[  -(\bar{z}-x)^{2}/2\right]  .$
Making this substitution and writing $\partial/\partial\bar{z}_{k}%
=(1/2)(\partial/\partial\tilde{x}_{k}+i\partial/\partial\tilde{y}_{k})$ we may
integrate by parts to get
\begin{align}
\frac{\partial f_{R}}{\partial x}  &  =ce^{x^{2}/p}\int_{\left|  \tilde{x}%
_{l}\right|  \leq R}\int_{\left|  \tilde{y}_{l}\right|  \leq R}e^{-\left(
\bar{z}-x\right)  ^{2}/2}\cdot\nonumber\\
&  \cdot F\left(  z\right)  \frac{\partial\left(  e^{-\tilde{x}^{2}%
/(p-1)}e^{-\tilde{y}^{2}}\right)  /\partial\bar{z}_{k}}{e^{-\tilde{x}%
^{2}/(p-1)}e^{-\tilde{y}^{2}}}e^{-\tilde{x}^{2}/(p-1)}e^{-\tilde{y}^{2}%
}d\tilde{x}\,d\tilde{y}\label{1diff}\\
&  +\frac{2x_{k}}{p}f_{R}\left(  x\right)  +\text{ boundary terms.}\nonumber
\end{align}
Note that we do not get a term involving the derivative of $F,$ since
$\partial F/\partial\bar{z}_{k}=0.$ Note also that the factor just after
$F\left(  z\right)  $ is linear in $\tilde{x}$ and $\tilde{y}.$ There will be
four boundary terms, in each of which the integrand will be $e^{-\left(
\bar{z}-x\right)  ^{2}/2}F\left(  z\right)  e^{-\tilde{x}^{2}/(p-1)}%
e^{-\tilde{y}^{2}}.$ Two terms will have integration over all of the variables
except $x_{k},$ with $x_{k}$ evaluated at $\pm R,$ and the other two will be
similar with the roles of $x_{k}$ and $y_{k}$ reversed. Our estimates on $F$
are sufficient to show that these boundary terms go to zero as $R$ tends to
infinity. So $\lim_{R\rightarrow\infty}\partial f_{R}/\partial x_{k}$ exists,
and the limit is easily seen to be locally uniform in $x.$ Since also
$\lim_{R\rightarrow\infty}f_{R}=f$, we conclude that $f$ is differentiable and
that
\begin{align}
\frac{\partial f}{\partial x_{k}}  &  =ce^{x^{2}/p}\int_{\mathbb{R}^{d}}%
\int_{\mathbb{R}^{d}}e^{-\left(  \bar{z}-x\right)  ^{2}/2}\cdot\nonumber\\
&  \cdot F\left(  z\right)  \frac{\partial\left(  e^{-\tilde{x}^{2}%
/(p-1)}e^{-\tilde{y}^{2}}\right)  /\partial\bar{z}_{k}}{e^{-\tilde{x}%
^{2}/(p-1)}e^{-\tilde{y}^{2}}}e^{-\tilde{x}^{2}/(p-1)}e^{-\tilde{y}^{2}%
}d\tilde{x}\,d\tilde{y}+\frac{2x_{k}}{p}f\left(  x\right)  .\label{2diff}%
\end{align}

Now we have already noted that $x^{\alpha}f\left(  x\right)  $ is in
$L^{p}(\mathbb{R}^{d},\rho).$ Since the factor after $F\left(  z\right)  $
grows only linearly, our estimates on $F$ will show that the first term is
also in $L^{p}(\mathbb{R}^{d},\rho).$ Thus $\partial f/\partial x_{k}$ is in
$L^{p}(\mathbb{R}^{d},\rho)$ as well. But now we may start our analysis anew
with the function $f$ replaced by $\partial f/\partial x_{k}$ and with the
inversion formula (\ref{inverse}) replaced by (\ref{2diff}). A similar
argument will show that $x^{\alpha}\left(  \partial f/\partial x_{k}\right)
\in L^{p}(\mathbb{R}^{d},\rho)$ for all $\alpha,$ and then that $x^{\alpha
}\left(  \partial^{2}f/\partial x_{l}\partial x_{k}\right)  \in L^{p}%
(\mathbb{R}^{d},\rho).$ We may keep on differentiating repeatedly, at each
stage getting a sum of terms of the form
\[
p_{1}\left(  x\right)  e^{x^{2}/p}\int_{\mathbb{R}^{d}}\int_{\mathbb{R}^{d}%
}e^{-\left(  \bar{z}-x\right)  ^{2}/2}F\left(  z\right)  p_{2}(\tilde
{x},\tilde{y})e^{-\tilde{x}^{2}/(p-1)}e^{-\tilde{y}^{2}}d\tilde{x}\,d\tilde{y}%
\]
where $p_{1}$ and $p_{2}$ are polynomials. Our estimates on $F$ are then
sufficient to give the required bounds on $(\partial/\partial x)^{\alpha}f. $
$\square$

Since the Segal-Bargmann transform $S$ is given in (\ref{sdef1}) as a
convolution, if $f$ satisfies Condition 2 then we will have $S\left[
(\partial/\partial x)^{\beta}f\right]  =(\partial/\partial z)^{\beta}F$ for
all $\beta.$ But $(\partial/\partial x)^{\beta}f$ again satisfies Condition 2,
and therefore $(\partial/\partial z)^{\beta}F$ satisfies Condition 1. Thus we
see that if $F$ satisfies Condition 1 then so do all the derivatives of $F.$

\subsection{Dimension-independent bounds}

\textit{Proof of Theorem \ref{hyper1.thm}}. We consider the number operator
$N$ defined on (a suitable dense subspace of) $L^{p}(\mathbb{R}^{d},\rho)$ by
\[
N=\sum_{j=1}^{d}\left[  -\frac{\partial^{2}}{\partial x_{j}^{2}}+x_{j}%
\frac{\partial}{\partial x_{j}}\right]  .
\]
We consider also the associated semigroup $\exp(-tN),$ which is a contraction
semigroup on $L^{p}(\mathbb{R}^{d},\rho)$ for all $1<p<\infty.$ We now
consider $p<2.$ Nelson's hypercontractivity theorem \cite{N} says that for all
$t\geq-\frac{1}{2}\log(p-1)$ the semigroup $\exp(-tN)$ maps $L^{p}%
(\mathbb{R}^{d},\rho)$ into $L^{2}(\mathbb{R}^{d},\rho),$ and
\[
\left\|  e^{-tN}f\right\|  _{L^{2}(\mathbb{R}^{d},\rho)}\leq\left\|
f\right\|  _{L^{p}(\mathbb{R}^{d},\rho)}.
\]

Meanwhile, \cite[Sect. 3e]{B1} (adapted to our normalization of the
Segal-Bargmann transform) tells us that
\[
S(e^{-tN}f)(z)=Sf(e^{-t}z).
\]
(More precisely, \cite{B1} establishes this if $f\in L^{2}(\mathbb{R}^{d}%
,\rho),$ but using the $L^{p}$ continuity properties of $S$ it is easily
established for $f\in L^{p},$ $p>1.$) Using the $L^{2}$ isometry property of
$S$ and hypercontractivity we conclude that
\[
\left\|  Sf(e^{-t}z)\right\|  _{L^{2}(\mathbb{C}^{d},\mu)}=\left\|
e^{-tN}f\right\|  _{L^{2}(\mathbb{R}^{d},\rho)}\leq\left\|  f\right\|
_{L^{p}(\mathbb{R}^{d},\rho)}%
\]
for all $t\geq-\frac{1}{2}\log(p-1).$ If we take $t=-\frac{1}{2}\log(p-1), $
then $e^{-t}=\sqrt{p-1}$ and we conclude that
\[
\left\|  Sf(\sqrt{p-1}z)\right\|  _{L^{2}(\mathbb{C}^{d},\mu)}\leq\left\|
f\right\|  _{L^{p}(\mathbb{R}^{d},\rho)}.
\]
This is the content of Theorem \ref{hyper1.thm}.$~\square$

\textit{Proof of Theorem \ref{hyper2.thm}}. Suppose $t$ is some positive
number. Suppose $F$ is a holomorphic function such that the function $F_{t}$
defined by
\[
F_{t}(z)=F(e^{t}z)
\]
is in $L^{2}(\mathbb{C}^{d},\mu).$ Define $f_{t}=S^{-1}F_{t}$ and $f=S^{-1}F.$
Again by \cite[Sect. 3e]{B1} we have
\[
f=e^{-tN}f_{t}.
\]
Hypercontractivity then implies that for all $p$ such that $t\geq\frac{1}%
{2}\log(p-1)$ we have $f\in L^{p}(\mathbb{R}^{d},\rho)$ and
\[
\left\|  f\right\|  _{L^{p}(\mathbb{R}^{d},\rho)}\leq\left\|  f_{t}\right\|
_{L^{2}(\mathbb{R}^{d},\rho)}.
\]
Thus
\[
\left\|  f\right\|  _{L^{p}(\mathbb{R}^{d},\rho)}\leq\left\|  f_{t}\right\|
_{L^{2}(\mathbb{R}^{d},\rho)}=\left\|  F_{t}\right\|  _{L^{2}(\mathbb{C}%
^{d},\mu).}%
\]

Now fix some $p>2$ and take $t=\frac{1}{2}\log(p-1)$, so that $e^{t}%
=\sqrt{p-1}.$ We conclude that
\[
\left\|  S^{-1}F\right\|  _{L^{p}(\mathbb{R}^{d},\rho)}\leq\left\|
S^{-1}F_{t}\right\|  _{L^{2}(\mathbb{R}^{d},\rho)}=\left\|  F_{t}\right\|
\]
where (with this value of $t$)
\[
\left\|  F_{t}\right\|  ^{2}=\int_{\mathbb{C}^{d}}\left|  F\left(  \sqrt
{p-1}z\right)  \right|  ^{2}\frac{e^{-\left|  z\right|  ^{2}}}{\pi^{d}}\,dz.
\]
This is the content of Theorem \ref{hyper2.thm}.$~\square$

\section{A primer on the Segal-Bargmann transform}

In this section I explain some of the whys and wherefores of the
Segal-Bargmann transform, specifically: the connection with the windowed
Fourier transform, the connection with coherent states, and the motivation for
using Gaussian measure on the domain space.

The Segal-Bargmann transform, the windowed Fourier transform, and coherent
states are all parts of ``harmonic analysis in phase space.'' (I have taken
this phrase from the title of the excellent book \cite{F} by Gerald Folland.)
On the one hand, these objects in the simplest case are equivalent to one
another, so that the same results have often been proved several times in
different settings. On the other hand, the types of generalizations considered
by the different groups are quite different.

Broadly speaking, harmonic analysis in phase space is the attempt to combine
information about a function and its Fourier transform into a single object.
In signal processing one considers a function of time and the associated phase
space is the plane, thought of as time-frequency space. In quantum mechanics
one considers a ``wave function'' on (say) $\mathbb{R}^{d}$ and the associated
phase space is $\mathbb{R}^{2d},$ thought of as position-momentum space. In
either case instead of considering the Fourier transform of a function $f$ in
its entirety one can consider ``the Fourier transform of $f$ near a point
$x$,'' which will then be a function of both $x $ and the frequency variable.
Both the Segal-Bargmann transform and the windowed Fourier transform give a
way of making sense out of this. In contrast to the wavelet expansion, both
the Segal-Bargmann and windowed Fourier transforms give information in a
\textit{fixed} distance scale around the point $x.$ The distinctive feature of
the Segal-Bargmann transform is that it is brings in holomorphic methods,
viewing $\mathbb{R}^{2d}$ as the complex space $\mathbb{C}^{d}.$

\subsection{Windowed Fourier transforms and holomorphic convolution transforms}

In the simplest (Gaussian) case the Segal-Bargmann transform and the windowed
Fourier transform are essentially the same thing. However there is a
difference in perspective that can be illustrated by considering two families
of transforms that have the Gaussian transform as their unique point of intersection.

First consider the general windowed Fourier transform. Given a Schwartz
function $\phi$ we consider a transform $\mathcal{F}_{\phi}$ given by
\[
\mathcal{F}_{\phi}\left(  f\right)  \left(  a,b\right)  =\int_{\mathbb{R}^{n}%
}f\left(  x\right)  \phi\left(  x-a\right)  e^{ib\cdot x}\,dx.
\]
If for example $\phi$ is compactly supported near the origin then
$\mathcal{F}_{\phi}$ is the Fourier transform of $f$ in a ``window'' near
$x=a.$ It is not hard to show, using the Plancherel theorem, that
$\mathcal{F}_{\phi}$ is (up to a constant) an isometry of $L^{2}%
(\mathbb{R}^{d})$ into $L^{2}\mathbb{(}\mathbb{R}^{2d}).$ The image may be
characterized by a certain reproducing kernel condition. See \cite[Sect.
3.1]{F} or \cite[p. 8]{FS}.

Meanwhile, consider ``holomorphic convolution transforms'' as follows. Let
$\nu$ be a strictly positive continuous function on $\mathbb{R}^{d}$ with
faster-than-exponential decay at infinity. Let
\[
\sigma\left(  k\right)  =\int_{\mathbb{R}^{d}}e^{2k\cdot b}\nu\left(
b\right)  \,db
\]
and let
\[
\phi\left(  z\right)  =\frac{1}{2\pi}\int_{\mathbb{R}^{d}}\frac{e^{-ik\cdot
z}}{\sqrt{\sigma\left(  k\right)  }}\,dk\quad z\in\mathbb{C}^{d}.
\]
Then $\phi$ is an entire holomorphic function on $\mathbb{C}^{d}$ \cite[Sect.
10]{H1}.

Given a function $\phi$ constructed in this way, define a transform $C_{\phi}
$ by
\[
C_{\phi}\left(  f\right)  \left(  z\right)  =\int_{\mathbb{R}^{d}}\phi\left(
z-x\right)  f\left(  x\right)  \,dx,\quad z\in\mathbb{C}^{d}.
\]
Then, again using just the Plancherel theorem, one can show \cite[Thm. 8]{H1}
that $C_{\phi}$ is a unitary map of $L^{2}(\mathbb{R}^{d})$ onto the space of
holomorphic functions on $\mathbb{C}^{d}$ that are square-integrable with
respect to the measure $\nu\left(  b\right)  \,da\,db,$ where $z=a+ib.$ One
can modify the definition to
\[
\tilde{C}_{\phi}\left(  f\right)  \left(  a+ib\right)  =\sqrt{\nu\left(
b\right)  }C_{\phi}\left(  f\right)  \left(  a+ib\right)  .
\]
Then $\tilde{C}_{\phi}$ maps $L^{2}(\mathbb{R}^{d})$ isometrically into
$L^{2}(\mathbb{R}^{2d}),$ and image of $\tilde{C}_{\phi}$ is precisely the set
of functions of the form $F\left(  z\right)  \sqrt{\nu\left(  b\right)  },$
where $F $ is holomorphic.

If we take $\nu\left(  b\right)  =\pi^{-d/2}\exp(-b^{2})$ then $\phi\left(
z\right)  =\left(  2\pi\right)  ^{-d/2}\exp\left(  -z^{2}/2\right)  $ and
\[
\tilde{C}_{\phi}\left(  f\right)  \left(  a+ib\right)  =e^{-ia\cdot b}%
\int_{\mathbb{R}^{d}}f\left(  x\right)  e^{-\left(  x-a\right)  ^{2}%
/2}e^{ib\cdot x}\,dx.
\]
In this case, except for the ``postmultiplier'' $e^{-ia\cdot b},$ we have just
a windowed Fourier transform with a Gaussian window. The map $C_{\phi}, $ with
$\phi$ Gaussian, is a form of the Segal-Bargmann transform, normalized in a
somewhat different way from the transform $S$ discussed in the rest of the
paper. (The two forms differ simply by a change of measure on the domain side
and the range side. On the domain side the change of measure is the ground
state transformation discussed below. On the range side the change of measure
is the map $F\left(  z\right)  \rightarrow c\,e^{z^{2}/4}F(z/\sqrt{2})$ which
brings about a change from the fully Gaussian measure $\mu$ used with the $S$
version of the transform to the partly Gaussian measure $\pi^{-d/2}\exp
(-b^{2})\,da\,db$ used with the $C_{\phi}$ version of the transform. See
\cite[Sects. 6.2 and 6.3]{H4} for more details.)

Note that in both the windowed Fourier transform and the holomorphic
convolution transform we are integrating a function $f$ against a family of
functions obtained from a single function $\phi.$ In the case of the windowed
Fourier transform these functions are obtained from $\phi$ by translations in
position and in frequency. In the case of the holomorphic convolution
transform the functions are obtained from $\phi$ by translations in position
and analytic continuation. It is not hard to show that one gets equivalent
results precisely if $\phi$ is a Gaussian. That is, suppose $\phi$ is a
Schwartz function on $\mathbb{R}^{d}$ that admits an entire analytic
continuation to $\mathbb{C}^{d},$ with moderate growth in the imaginary
directions. Suppose also that there exist constants $c_{a,b}$ such that
\[
\phi\left(  x-(a+ib)\right)  =c_{a,b}\phi\left(  x-a\right)  e^{ib\cdot x}%
\]
for all $a,b\in\mathbb{R}^{d}.$ Then necessarily $\phi$ is of the form
\[
\phi\left(  x\right)  =\alpha e^{-x^{2}/2}e^{i\beta\cdot x}%
\]
for some $\alpha,\beta.$ Thus the windowed Fourier transform can be re-cast in
holomorphic terms only in the Gaussian case.

\subsection{Coherent states}

The transforms in the previous subsection consist of integrating a function
$f$ against a family of functions parameterized by points $\left(  a,b\right)
$ in the phase space $\mathbb{R}^{2d}.$ These functions are typically
localized in position near $a$ and in frequency near $b.$ Functions of this
sort are often called \textit{coherent states} in the physics literature, with
the name suggesting the phase space localization. The Gaussian wave packets
$e^{-\left(  x-a\right)  ^{2}/2}e^{ib\cdot x}$ are the \textit{canonical
coherent states} studied by Schr\"{o}dinger and von Neumann in the early days
of quantum mechanics (though without using that name). These states behave in
a very classical way under the time-evolution of a quantum harmonic
oscillator. In general coherent states in physics are thought of as quantum
states that approximate as well as possible a classical particle with a fixed
position $a$ and momentum $b.$

Many generalizations of the canonical coherent states have been considered;
see for example the book \cite{KS}. The generalizations often involve geometry
or Lie group representation theory. For example, Perelomov \cite{P} replaces
the Heisenberg group action (i.e. translations in position and in frequency)
by an irreducible unitary representation $\pi$ of a Lie group $G$ acting on a
Hilbert space $H.$ The coherent states are then the vectors of the form
$\pi\left(  g\right)  \phi_{0},$ where $\phi_{0}$ is a fixed non-zero vector
in $H.$ In a different direction, J. Rawnsley, M. Cahen, and S. Gutt, building
on ideas of F. Berezin, have considered ``quantization'' of K\"{a}hler
manifolds $M$ (e.g. \cite{RCG}). In this case one builds a Segal-Bargmann type
Hilbert space of holomorphic functions over $M,$ and the coherent states are
the elements $\chi_{m}$ of the Segal-Bargmann space such that $\left\langle
\chi_{m},F\right\rangle =F\left(  m\right)  ,$ for $m\in M$ and $F$ in the
Segal-Bargmann space over $M.$ The Perelomov and K\"{a}hler constructions
intersect in the case of homogeneous K\"{a}hler manifolds, yielding for
example the Bott-Borel-Weil construction of the irreducible representations of
compact semisimple Lie groups. Still another type of generalized coherent
states has been considered by the author, namely, coherent states on compact
Lie groups defined in terms of heat kernels. See \cite{H1,H5}.

Typically in the physics literature it is assumed that one has a ``resolution
of the identity'' for the coherent states \cite{KS}. Suppose that the coherent
states are a family of vectors $\psi_{\alpha}$ in a Hilbert space $H,$
parameterized by points $\alpha$ in some parameter space $P.$ Then a
resolution of the identity means a decomposition of the identity operator on
$H$ as
\begin{equation}
I=\int_{P}\left|  \psi_{\alpha}\right\rangle \left\langle \psi_{\alpha
}\right|  \,d\mu\left(  \alpha\right) \label{res.id}%
\end{equation}
for some measure $\mu$ on $P.$ Here $\left|  \psi_{\alpha}\right\rangle
\left\langle \psi_{\alpha}\right|  $ means the orthogonal projection onto the
vector $\psi_{\alpha}.$

Given such a resolution of the identity we have
\[
v=\int_{P}\psi_{\alpha}\left\langle \psi_{\alpha},v\right\rangle \,d\mu\left(
\alpha\right)
\]
for all $v\in H.$ That is, each $v\in H$ has a preferred (but not unique!)
expansion in terms of the coherent states, the coefficients of which are
computed simply by taking the inner product of $v$ with each $\psi_{\alpha}.$
Furthermore we have
\begin{align*}
\left\langle v,v\right\rangle  &  =\int_{P}\left\langle v,\psi_{\alpha
}\right\rangle \left\langle \psi_{\alpha},v\right\rangle \,d\mu\left(
\alpha\right) \\
&  =\int_{P}\left|  \left\langle \psi_{\alpha},v\right\rangle \right|
^{2}\,d\mu\left(  \alpha\right)  .
\end{align*}
This means that we can define a generalized Segal-Bargmann transform
$S:H\rightarrow L^{2}(P,\mu)$ by $S\left(  v\right)  \left(  \alpha\right)
=\left\langle \psi_{\alpha},v\right\rangle ,$ and $S$ will be an isometry of
$H$ into (but not onto) $L^{2}(P,\mu).$

Conversely, if one has an isometric transform of Segal-Bargmann type, one can
reformulate the isometricity as a resolution of the identity. So at some level
Segal-Bargmann transforms are equivalent to coherent states. Nevertheless
there is a different emphasis in the two settings.

\subsection{The ground state transformation}

One special feature of the Segal-Bargmann transform is that it is commonly
normalized so that its domain is an $L^{2}$ space over $\mathbb{R}^{d}$ with
respect to a Gaussian measure. The transformation from Lebesgue measure to
Gaussian measure is a special case of the \textit{ground state transformation}%
, which has a long and distinguished history in mathematical physics. (See for
example \cite[Sect. 4]{G}, \cite{DS}, or \cite[Sects. 3.2-3.4]{GJ}.) The
Gaussian wave packets (or coherent states) $\exp[-\left(  z-x\right)  ^{2}/2]$
are evidently obtained from the function $\phi_{0}\left(  x\right)
:=\exp[-x^{2}/2]$ by complex translations. The function $\phi_{0}$ is the
ground state of a quantum harmonic oscillator. This means that
\[
-\Delta\phi_{0}\left(  x\right)  +x^{2}\phi_{0}\left(  x\right)  =d\phi
_{0}\left(  x\right)  ,
\]
where $d$ is the minimum of the spectrum of the operator $-\Delta+x^{2}$
acting in $L^{2}(\mathbb{R}^{d},dx).$ In studying this operator it is
convenient to divide all functions by the ground state, thus considering the
map $f\rightarrow f/\phi_{0},$ which is a unitary map of $L^{2}(\mathbb{R}%
^{d},dx)$ onto $L^{2}(\mathbb{R}^{d},\phi_{0}^{2}\left(  x\right)  \,dx).$

More generally the operation of dividing by the ground state arises naturally
in the path-integral formulation of quantum mechanics (not just for the
harmonic oscillator), and is necessary if one wants to let the dimension $d$
tend to infinity. This infinite-dimensional limit is what arises in quantum
field theory, and indeed Segal \cite{S} worked in the $d=\infty$ setting from
the beginning.

Another reason for the change from Lebesgue to Gaussian measure is that it is
in the setting of Gaussian measures that the story of hypercontractivity and
logarithmic Sobolev inequalities gets told. (See \cite{DGS} for a summary of
this story.) In particular, the scale of $L^{p}$ spaces with respect to
Gaussian as opposed to Lebesgue measure is extremely important. See, for
example, the last part of Section 2 for an application of hypercontractivity.

This brings us to the Gaussian measure formulation of the Segal-Bargmann
transform $S,$ which is a unitary map of $L^{2}(\mathbb{R}^{d},\rho)$ onto
$\mathcal{H}L^{2}(\mathbb{C}^{d},\mu).$ In this form the measure on
$\mathbb{R}^{d}$ is Gaussian and the measure on $\mathbb{C}^{d}$ is Gaussian
in both the real and imaginary directions. This form differs from that
considered by Bargmann in \cite{B1} simply by the ground state transformation.
It is natural to consider the image of $L^{p}(\mathbb{R}^{d},\rho)$ under $S,$
and this is the subject of the present paper.

\end{document}